# Magnetic and Thermodynamic properties of face-centered cubic Fe-Ni alloys

M.Yu. Lavrentiev [*], J.S. Wrobel, D. Nguyen-Manh, and S.L. Dudarev

*CCFE, Culham Science Centre, Abingdon, Oxon, OX14 3DB, UK*

[*] Corresponding Author, E-mail: Mikhail.Lavrentiev@ccfe.ac.uk

**Abstract**

A model lattice *ab initio* parameterised Hamiltonian spanning a broad range of alloy compositions and a large variety of chemical and magnetic configurations has been developed for face-centered cubic Fe-Ni alloys. Thermodynamic and magnetic properties of the alloys are explored using configuration and magnetic Monte Carlo simulations in a temperature range extending well over 1000 K. The predicted face-centered cubic – body-centered cubic coexistence curve, the phase stability of ordered $Fe_3Ni$, $FeNi$, and $FeNi_3$ intermetallic compounds, and the predicted temperatures of magnetic transitions simulated as functions of alloy composition agree well with experimental observations. Simulations show that magnetic interactions stabilize the face-centered cubic phase of Fe-Ni alloys. Both the model Hamiltonian simulations and *ab initio* data exhibit a particularly large number of magnetic configurations in a relatively narrow range of alloy compositions corresponding to the occurrence of the Invar effect.

# 1. Introduction

Developing structural materials that retain their engineering properties over extended periods of time at high temperature and high radiation dose is one of the major challenges for fusion and fission materials science and technology. Developing tools for modelling such materials, particularly steels, is one of the objectives for the European fusion programme [1]. Face-centered cubic (fcc) Fe-Ni-Cr based austenitic steels retain good engineering strength at high temperatures, making them attractive candidate materials for fusion and advanced fission technology. In particular, austenitic 304 and 316 steels are used as structural materials for light water and fast breeder fission reactors. These steels contain about 10at. % Ni and 20 at. % Cr [2,3], so that any methodology for modelling such steels should in the first place be able to model ternary Fe-Ni-Cr alloys in the composition range where neither of the constituent elements can be treated as impurity. An extra factor that must be taken into account when modelling iron-based alloys and steels is the fact that the phase stability of iron-based alloys is controlled by magnetism, for example magnetism stabilizes the ferritic body-centered cubic (bcc) phase of pure iron under ambient conditions. Previous theoretical works on the phase diagram of Fe-Ni system were based on the interatomic interaction potentials [4-7] and so were unable to describe magnetic phase transitions in that system.

Magnetic Cluster Expansion (MCE) simulations of Fe-Cr alloys [8-11] showed that a model based on an *ab initio* parameterized MCE Hamiltonian was able to describe a broad range of magnetic and structural transformation effects in bcc iron-based magnetic alloys. This has prompted us to develop an MCE parameterization for face-centered cubic Fe-Ni alloys as a step towards the treatment of ternary Fe-Ni-Cr alloys. The phase diagram of binary Fe-Ni alloys [12] shows that the solubility of Ni in bcc iron is very low, as opposite to the high solubility of Ni in fcc Fe, where the two species are fully soluble in the entire range of concentrations at high temperature. Hence a realistic MCE Hamiltonian should be expected to be able to describe the miscibility gap between fcc Fe-Ni, bcc Fe and bcc Fe-Ni. Another important feature of the phase diagram is the presence of an ordered $L1_2$ structure with $FeNi_3$ composition. Also, the possible occurrence of ordered FeNi and $Fe_3Ni$ compounds (the first of them was found in meteorites [13-15]) should be explored.

This paper is organized as follows. In Section 2, we parameterize the MCE Hamiltonian for the fcc Fe-Ni alloy. Magnetic properties of pure fcc iron and nickel and their solid solutions are investigated in Section 3, and the phase diagram of the system is explored in Section 4. We summarize the results of our study and conclude in Section 5.

## 2. Parameterization of Magnetic Cluster Expansion

The main principles of MCE are described in Refs. [8,9,16]. The general functional form of the Heisenberg-Landau Hamiltonian used in MCE simulations is:

$$H(\{\sigma_i\},\{\mathbf{M}_i\}) = NI^{(0)} + I^{(1)}\sum_i \sigma_i + \sum_{ij \in 1NN} I_{ij}^{(1NN)}\sigma_i\sigma_j + \sum_{ij \in 2NN} I_{ij}^{(2NN)}\sigma_i\sigma_j + ...$$
$$+ \sum_i \left( A^{(0)} + A^{(1)}\sigma_i + \sigma_i \sum_j A_{ij}^{(2)}\sigma_j \right)\mathbf{M}_i^2 + \sum_i \left( B^{(0)} + B^{(1)}\sigma_i + \sigma_i \sum_j B_{ij}^{(2)}\sigma_j \right)\mathbf{M}_i^4 + ...$$
$$+ \sum_{ij \in 1NN} J_{ij}^{(1NN)}\mathbf{M}_i \cdot \mathbf{M}_j + \sum_{ij \in 2NN} J_{ij}^{(2NN)}\mathbf{M}_i \cdot \mathbf{M}_j + ... \qquad (1)$$

Here, $I$'s are the non-magnetic Cluster Expansion (CE) coefficients; parameters $A$ and $B$ represent the configuration-dependent Landau coefficients for the magnetic self-energy terms, and $J$'s are the inter-lattice-site Heisenberg magnetic interaction parameters. The functional form of the MCE Heisenberg-Landau Hamiltonian (1) guarantees that the magnetic self-energy terms, and hence the directions *and* magnitudes of atomic magnetic moments $\mathbf{M}_i$ predicted by the model, depend on the local environment of each atom in the alloy. Hamiltonian (1) is based on the undistorted rigid lattice approximation, which is valid for Fe-Ni alloys where the atomic radii of the two elements are similar (see, e.g., Table 1 in [17]). We note that the functional form of Hamiltonian (1) can be extended and generalized further. In particular, the Landau expansion for magnetic self-energy can be extended beyond the quadratic and quartic terms. Below we show that such extension is actually necessary for the treatment of magnetic properties of fcc Fe-Ni alloys.

Whereas Hamiltonian (1) can be parameterized in several different ways, still the search for a suitable set of Landau and Heisenberg parameters always begins by selecting a set of structures, the energies and magnetic moments of which are calculated *ab initio*. In the fcc FeNi alloy case, a set of 29 alloy configurations was investigated by *ab initio* DFT calculations, together with the two limiting cases of pure fcc iron and fcc nickel. Spin-polarized DFT calculations were performed using the projector augmented wave (PAW) method [18] implemented in the VASP package [19-21]. Exchange-correlation was treated within the generalized gradient approximation

GGA-PBE [22]. The total energies and magnetic moments were calculated assuming the plane-wave cutoff energy of 400 eV and a k-point mesh with spacing of 0.26 Å$^{-1}$.

An *ab initio* investigation of ferromagnetic fcc iron shows the occurrence of the two so-called high-spin and low-spin magnetic states. The high-spin configuration is found to be slightly more energetically favourable than the low-spin one, but the difference is of the order of 10 meV/atom or smaller, resulting in that both configurations contribute equally significantly to the finite temperature magnetic properties of the alloy. The occurrence of high- and low-spin magnetic configurations in the ground state of the alloy cannot be accounted for by a Landau-type Hamiltonian involving only the quadratic and quartic terms, since such functional form of the Hamiltonian exhibits only one minimum as a function of the *magnitude* of magnetic moment vector. In order to describe both the high- and low-spin magnetic configurations of fcc Fe, it is necessary to include terms up to the 8$^{th}$ order in magnetic moment in the Landau expansion. We use DFT data to fit the energy of pure fcc ferromagnetic Fe as a function of magnetic moment. Figure 1 shows that an 8$^{th}$ order magnetic moment Landau expansion agrees well with *ab initio* results, replicating the correct difference between the energies of low-spin and high-spin magnetic configurations.

In an earlier MCE study of bcc-fcc transitions in Fe and Fe-Cr alloys, we showed that extending the range of Heisenberg interaction parameters to the third nearest neighbour was sufficient for modelling magnetic configurations in pure fcc iron [9]. When fitting the Heisenberg parameters for the Fe-Ni system, we found that it was necessary to extend the range up to the fourth nearest neighbour in order to describe the Ni-Ni and Ni-Fe interactions. Hence to retain consistency we decided to extend the range of magnetic and non-magnetic parameters to the 4$^{th}$ nearest neighbour for Fe-Fe interactions as well. For each of the 29 alloy configurations, as well as for pure nickel and iron, the *ab initio* energy data (including, for pure iron, several ferro- and antiferromagnetic configurations) were used to evaluate a trial set of MCE parameters. Also, the derivatives of MCE energy with respect to magnetic moments were calculated numerically. The sum of squares of deviations of MCE predictions from the DFT data,

$$S = \sum_i \alpha_i \left(E_i^{DFT} - E_i^{MCE}\right)^2 + \sum_i \beta_i \sum_j \left(\frac{\partial E_i^{MCE}}{\partial \mathbf{M}_j}\right)^2 \qquad (2)$$

was taken as a measure of goodness of MCE fit. Here, we include both the differences between the MCE and DFT energies, and also a measure of deviation of the position of the energy minimum predicted by MCE from its DFT minimum value, characterized by the sum of squares of derivatives of the MCE energy with respect to atomic magnetic moments. Coefficients $\alpha_i$ and $\beta_i$ were adjusted during the fitting procedure to reflect contributions of various configurations to the fit and are, in general, chosen to be larger for the lower energy configurations. The resulting values of magnetic Heisenberg parameters are given in Table 1. DFT and MCE energies of mixing for the structures used in the fit are compared in Figure 2. The mean square deviation of *ab initio* energies from those predicted by MCE is 12 meV. For the non-magnetic interaction parameters, only the two-atom clusters were considered, and the values of parameters $I_{ij}$ are -6.650 meV, 3.451 meV, 0.104 meV, and -0.580 meV for the 1$^{st}$, 2$^{nd}$, 3$^{rd}$ and 4$^{th}$ nearest neighbours, respectively.

Monte Carlo simulations of pure Fe and Ni and Fe-Ni alloys were performed as follows. The simulation box contained 16384 atoms (16×16×16 face centered cubic unit cells each containing 4 atoms). For the chemically ordered structures, at each Monte Carlo step a trial random change of magnetic moment of a randomly chosen atom was attempted and accepted or rejected according to the Metropolis criterion. Both the thermalization and accumulation stages included 40000 attempts per atom. For the low temperature cases and complex magnetic structures of pure iron, simulations involving 128000 attempts per atom were also performed to ensure that the system has reached equilibrium. For the case of random structures, we used two types of simulations. The first was similar to the one used for the ordered alloys, but with random configurations corresponding to a given Fe and Ni content. In comparison with *ab initio* calculations, where relatively small size of the simulation box requires using special quasi-random structures (SQS) (see, e.g., [23]), large-scale Monte Carlo simulations can be performed by simply choosing completely random configurations for the two types of atoms. In order to verify the results, several comparisons with other random structures were made by changing the seed of the random number generator. Thermodynamic data obtained in this way correspond to a completely random system and are hence characterized by a known value of configurational entropy, namely

$$S_{conf} = k_B \left( x_{Fe} \ln(x_{Fe}) + (1 - x_{Fe}) \ln(1 - x_{Fe}) \right) \tag{3}$$

Another approach is the exchange Monte Carlo, in which trial changes of magnetic moment are combined with attempts to exchange two randomly chosen atoms of different species. This approach was previously successfully used in Monte Carlo simulations based on interatomic interaction potentials [24-27], as well as on non-magnetic cluster expansion [28]. It gives reliable results for the enthalpy of mixing, but the configurational entropy and thus the free energy of mixing is difficult to evaluate. By comparing the enthalpies of mixing obtained using the two approaches, we estimate how significantly a given configuration of the alloy deviates from a completely random mixture.

**3. Magnetic and Thermodynamic Properties**

3.1 Magnetism of pure Fe and Ni

For pure fcc iron, the first and second nearest-neighbour Heisenberg magnetic interaction parameters favour ferromagnetic ordering of moments, whereas the third and the fourth favour antiferromagnetic ordering (see Table 1). Interplay between those interactions results in that the lowest-energy magnetic configuration is non-collinear antiferromagnetic. The energies of several magnetic configurations of fcc iron are shown and compared with DFT results in Table 2. The predicted non-collinear magnetic ground state agrees with the findings of our MCE study of bcc-fcc transitions in Fe and Fe-Cr alloys [9], even though the set of MCE parameters used here is different from that used in Ref. [9]. Non-collinear antiferromagnetic ordering of magnetic moments was experimentally discovered at low temperatures in fcc Fe precipitates embedded in a Cu matrix [29, 30]. Non-collinear magnetism is realised in magnetic materials with competing magnetic coupling parameters, where magnetic interactions are comparable in terms of their magnitude and change sign as a function of interatomic distance. In the earlier study of fcc Fe-Cr, the nearest neighbour Fe-Fe interaction was taken as ferromagnetic, while the second and third nearest neighbour Fe-Fe interaction parameters were antiferromagnetic ([9], Table 1). In the current MCE Hamiltonian, extending the Landau expansion to the $8^{th}$ order in magnetic moment to model the high-spin and low-spin magnetic states resulted in the increase of the magnitude of the (negative, ferromagnetic) second order magnetic term. In our fit this corresponds to strongly ferromagnetic second nearest neighbour interactions. Although the antiferromagnetic $3^{rd}$ and $4^{th}$ nearest neighbour interaction parameters are smaller in terms of their magnitude (see Table 1), the relatively large number of the third (24 atoms) and fourth (12 atoms) nearest neighbours in fcc lattice are sufficient to overwhelm the effect of ferromagnetic interaction involving only six second nearest neighbours. As a result, the antiferromagnetically ordered single and double layer structures have energies lower than that of the ferromagnetic configuration. Still, the ground state antiferromagnetic collinear configuration of fcc Fe turns out to be magnetically frustrated, and frustration is partially resolved by rotating the magnetic moments away from collinearity. We observed similar behaviour of magnetic moments in both *ab initio* and MCE studies of Fe-Cr interfaces in bcc lattice [31].

While the total magnetic moment of the alloy is zero due to averaging over all the possible directions of atomic magnetic moments, the average magnitude of magnetic moment vector of an individual Fe atom $\langle |\mathbf{M}| \rangle = \frac{1}{N} \sum_{i=1}^{N} |\mathbf{M}_i|$ does not vanish. The slightly lower energy of the high-spin state compared to the low-spin one ensures that at low temperatures the average magnitude of atomic magnetic moment is of the order of 2.648 $\mu_B$. As temperature increases, the slope of energy variation as a function of temperature changes at about 450 K (Figure 3a), indicating a transition from an antiferromagnetic to a paramagnetic state. This temperature is higher than that found using the parameterization given in Ref. [9], though the Néel temperature is still very low compared to the temperature range of stability of fcc iron (1185 K – 1667 K [32]). Below the magnetic transition, the average magnitude of an atomic magnetic moment decreases with temperature due to the mixing of high-spin and low-spin magnetic states, eventually decreasing to 2 $\mu_B$. Above the transition, where the moments are completely disordered, the average length of the magnetic moment $\langle |\mathbf{M}| \rangle$ increases slowly as a function of temperature, see Figure 3b.

For pure fcc nickel, Monte Carlo simulations predict strong collinear ferromagnetic ordering at low temperatures, in agreement with experiment and *ab initio* calculations. This can also be deduced from the set MCE parameters given in Table 1, where for almost all the neighbours the Heisenberg interaction parameters are ferromagnetic. The low temperature value of atomic magnetic moment found in MCE simulations, 0.575 $\mu_B$, is within 5 % margin of experimentally observed value of 0.605 $\mu_B$ [33]. Rising temperature destroys magnetic order at 550-600 K according to the magnetic moment data shown in Figure 4a. This agrees well with the experimental Curie temperature of nickel of 631 K [33], confirming the good accuracy of our MCE parameter set. The occurrence of magnetic phase transition just below 600 K is confirmed also by the peak in the temperature dependent magnetic part of the specific heat shown in Figure

4b.

3.2 Magnetism and the enthalpy of mixing of Fe-Ni alloys

Experimental and theoretical data on the enthalpy of mixing of Fe-Ni alloys are relatively scarce. A recent study by Idczak *et al*. [34] was performed using alloy samples with iron content below 10 at. % where the data were subsequently extrapolated to the entire range of alloy compositions. The results exhibit negative enthalpy of mixing with a minimum of approximately 70 meV/atom, whereas in the interatomic potential studies [6] a much smaller minimum value of about 40 meV/atom was found. Our results for the enthalpies of mixing, computed assuming random alloy configurations, are shown in Figure 5. The reference energies at the limits of the composition range (pure Fe and pure Ni) were taken as those of ground state magnetic configurations, i.e. non-collinear antiferromagnetic for Fe and ferromagnetic for Ni. The minimum value of the enthalpy of mixing is close to -100 meV/atom, which is lower than what is found in Ref. [34]. However, this prediction is in good agreement with our own DFT calculations, which exhibit even lower enthalpies of mixing for several ordered structures (see Figure 2). With increasing temperature the absolute value of the enthalpy of mixing increases slightly, as shown in Figure 5. Even lower negative enthalpies of mixing are found in simulations describing quenching. In such simulations, we perform exchanges between atoms and simultaneously decrease the temperature, allowing the system to find the lowest-energy configuration. Figure 6 shows the result of such quenching, where the enthalpies of mixing go as low as -160 meV/atom. The enthalpies of formation are of course much higher at the Fe end of the concentration range, because bcc iron has lower energy than fcc Fe. Our DFT calculations predict that the formation energy of ferromagnetic bcc Fe is 107 meV/atom lower than the lowest energy antiferromagnetic double layer configuration of fcc Fe.

When discussing the magnetic behaviour of the system, it is important to mention that the results differ between collinear and non-collinear MC simulations. While the *ab initio* calculations used for fitting MCE always assume collinear magnetic order, finite temperature MCE model treats both collinear and non-collinear magnetic configurations. As we already showed in Section 3.1, the ground state of pure fcc Fe is non-collinear antiferromagnetic. Relatively strong ferromagnetic interactions between Fe and Ni atoms in the second and third nearest neighbour positions result in that the magnetic configuration of random FeNi alloy approaches collinearity as the nickel content increases. The system becomes ferromagnetic and almost completely collinear. Because the magnetic moment of Fe is much larger than that of Ni, at some concentration the total magnetic moment of the system becomes higher than that of pure Ni, resulting in a maximum of magnetic moment as a function of Fe content shown in Figure 7. At elevated temperatures, magnetic disorder rapidly destroys ferromagnetism for almost all the iron concentrations (Figure 7), except for the very low Fe alloys, where nonvanishing magnetic moment survives up until the Curie transition temperature in nickel.

3.3 Magnetism of ordered Fe-Ni compounds

The ordered alloy compounds that we investigated in this paper are $FeNi_3$ with $L1_2$ structure, FeNi with $L1_0$ structure, and $Fe_3Ni$ with Z1 and $L1_2$ structures. The $L1_2$ $FeNi_3$ structure plays an important rôle in the phase diagram of Fe-Ni, being ferromagnetic up to 940 K according to experiment [35]. The FeNi with $L1_0$ structure (tetrataenite) is found in meteorites [13-15] and has the lowest energy of all the alloy configurations with the same stoichiometry, according to DFT calculations. For $Fe_3Ni$, until recently the $L1_2$ superlattice was believed to have the lowest energy among ordered structures [36]. However, it was found recently by Barabash *et al.* [37] that the Z1 superlattice is more stable than the $L1_2$ structure. In any case, the energies of these structures are fairly close and this prompted us to investigate both of them.

In the absence of constraint on the directions of magnetic moments, three out of the four structures studied here ($L1_0$ FeNi and both $FeNi_3$ and $Fe_3Ni$ with $L1_2$ structure) are ferromagnetic at low temperatures. In Figure 8a we show the temperature dependence of magnetic moment found for these three compounds. Similarly to the case of random Fe-Ni mixtures, the total magnetic moment of the system is the highest in FeNi. Magnetic order in all of them is almost exactly collinear. We note that in the case of $Fe_3Ni$, the magnitudes of magnetic moments on both Fe and Ni in L12 are substantially smaller than in pure metals and other ordered compounds. The average magnitude of magnetic moment of Fe, $\langle|\mathbf{M}|\rangle_{Fe}$, in $Fe_3Ni$ was equal to 1.880 $\mu_B$, as compared to 2.648 $\mu_B$ in pure Fe. This indicates that the low-spin state of iron can potentially be stabilized by small amounts of nickel. The magnetic moment of Ni, $\langle|\mathbf{M}|\rangle_{Ni}$, in $Fe_3Ni$ is 0.503 $\mu_B$, which is also smaller than in pure Ni (0.60 $\mu_B$ according to DFT calculations, 0.575 $\mu_B$ according to the MCE simulations) and in other ordered systems.

In our simulations the energy of the Z1 structure is found to be approximately 9 meV/atom higher than the energy of the $L1_2$ structure. This value is within the error-bar of values predicted by MCE simulations. The Z1 $Fe_3Ni$ superlattice is non-collinear antiferromagnetic, which indicates that its magnetic structure is related to that of the ground state of pure fcc Fe. The main structural difference between the Z1 and $L1_2$ structures is that in Z1, iron and nickel atoms are packed in planes, with Ni plane followed by three Fe planes. In the $L1_2$ structure, there are alternating Fe and Fe-Ni planes, so that the Z1 superlattice is more segregated, and contains larger volumes of pure Fe. In our opinion, this segregation of Fe and Ni is the main reason explaining the occurrence of non-collinear antiferromagnetism in Z1, whereas in the more mixed $L1_2$, as well as in random Fe-Ni mixture of the same composition, ferromagnetic order is prevalent. It is worth noting that if Monte Carlo simulations are restricted to collinear magnetism, the Z1 structure is also ferromagnetic at low temperatures, with the moments of Fe and Ni being parallel. The energy of a collinear ferromagnetic Z1 structure is 6.5 meV/atom higher than the energy of a non-collinear antiferromagnetic configuration. The average magnitude of magnetic moment of Fe equals 2.645 $\mu_B$, which is very close to that of pure fcc Fe. This agrees with our DFT results, which also predict the magnetic moment of iron in ferromagnetic Z1 structure to be greater than 2.5 $\mu_B$. We believe that antiferromagnetic order can be destroyed more easily by the relatively small additions of Ni in the case of collinear magnetism because the nearly collinear antiferromagnetic fcc Fe structures have energies closer to ferromagnetic fcc Fe than the non-collinear antiferromagnetic fcc Fe configurations.

Of the three ferromagnetic structures studied, temperature destroys the magnetic order first in $L1_2$ $Fe_3Ni$, where it vanishes already at 500 K (Figure 8a). At the same time, FeNi and $FeNi_3$ remain magnetically ordered until very high temperatures, ~1000 K and ~1200 K, respectively. For $FeNi_3$ the predicted Curie temperature is higher than the experimentally observed one. However, as was noted in [38], the experimental Curie point at those high temperatures is an underestimation because of fast self-diffusion and the resulting difficulties associated with maintaining the chemical order. For comparison, in the random mixture of $FeNi_3$, magnetic order vanishes already at 500-550 K whereas in a random mixture with FeNi composition, the corresponding temperature is 400-450 K.

Magnetic interaction between Fe and Ni results in both elements retaining magnetic order until the Curie point, as can be observed for ordered FeNi in Figure 8b. Magnetic order is retained at temperatures much higher than the Curie point in pure Ni for both nickel and iron constituents.

This behaviour is similar to that characterizing the magnetization of layers of chromium at the interface with Fe, where the magnetic moment of several atomic layers remains nonzero well above the Néel temperature of Cr [31].

**4. Phase diagram of Fe-Ni alloy**

Using the parameters described above, we performed Monte Carlo simulations to establish the equilibrium phase diagram at the interface between the bcc and fcc phases. The solubility of Ni in bcc iron is extremely low, with experimental phase diagram predicting approximately 5 at. % Ni solubility limit at about 800 K, and even smaller values at lower temperatures. Hence the free energy of bcc Fe-Ni system can be approximated by the free energy of pure bcc iron. Calculations of the free energy of bcc Fe and the difference between bcc and fcc Fe free energies were performed in our previous work where we predicted the occurrence of bcc-fcc phase transitions in Fe and Fe-Cr, corresponding to the γ-loop in the phase diagram [9]. Using those results and the new data for fcc Fe-Ni, the usual tangent construction can be applied to evaluate the coexistence curve. An example of such construction for T=600 K is given in Figure 9. As discussed in Section 2, we performed two separate calculations for the Fe-Ni system: one for a random mixture and another where atomic exchanges were included. We found that the free energy of a random mixture was always lower than the enthalpy of mixing of a system where atoms were allowed to exchange. This was observed for all the temperatures and for all the concentrations studied here. In order to provide the lower estimate for the free energy, we added the ideal configuration entropy (3) to the enthalpy of the alloy system where exchanges were permitted. The two tangent constructions gave almost identical results for the bcc-fcc coexistence curve, as can be seen in Figures 9 and 10. The predicted coexistence curve (Figure 10) shows that the area of the phase diagram where fcc Fe-Ni alloys are stable is very broad, in agreement with experimental data [12]. Note that our calculations do not take into account lattice vibrations. This approximation is well justified for the case of Fe-Ni alloy, where the vibrational entropy of alloying is very low. In a study of Bogdanoff and Fultz [39], based on the experimental information about the phonon density of states, the vibrational entropy of alloying was found to be $\Delta S_{vib}^{alloy} = 0.02\,k_B$/atom for the Fe$_3$Ni compound with L1$_2$ structure, which is smaller than the error-bar of ±0.03 $k_B$/atom, and much smaller than the ideal configurational entropy, which is 0.562 $k_B$/atom at that composition.

In the high-Ni part of the phase diagram, it is necessary to compare the energies of ordered structures ($L1_2$ for the case of $FeNi_3$ and $L1_0$ for the case of FeNi) with the free energies of random mixtures for the same alloy composition. For the ordered structures, the configuration entropy is zero. For the random mixtures, we again used expression (3) for the ideal configuration entropy. The free energies of ordered and random systems are compared in Figure 11. For FeNi, random structures become more energetically favourable at about 520 K, for the $FeNi_3$ compound – at about 730 K. These numbers should be compared with experimental temperatures of 593 K and 770 K, respectively [40,41]. Small underestimation of the order-disorder transition temperatures compared to experiment can be related to the fact that the actual configuration entropy is slightly higher than the ideal one because of the remaining order in the structures at low temperatures. Summarizing those results, we conclude that while the ordered $FeNi_3$ system is certainly stable until high enough temperatures, for the FeNi $L1_0$ structure the temperature of the order-disorder transformation is relatively low, meaning that it might take a very long time for the alloy of that composition to reach the ordered state during cooling. Combining transition temperatures for the ordered ferromagnetic and disordered antiferromagnetic systems (520 K and 730 K) with the Curie temperature of Ni, which our simulations predict to be 550-600 K, we are able to explain the experimentally observed maximum of magnetic transformation temperature found as a function of nickel content.

5. Discussion

The effect of magnetism on the thermodynamic properties of solids and structural phase transitions has long been recognized. For example, the ferromagnetic-paramagnetic transition in iron at 1043 K is responsible for the bcc-fcc transition at 1185 K [9,42,43]. Recent advances in simulation algorithms expanded the range of systems accessible to simulation from pure metals to binary alloy solutions with different degree of chemical disorder. In this paper, we show how Magnetic Cluster Expansion can be applied to magnetic fcc Fe-Ni alloy.

The main difference between the phase diagram of Fe-Cr studied earlier, and Fe-Ni alloys, is the fact that the area of the phase diagram where fcc structure is stable, is very large. The so-called γ-loop in the Fe-Cr phase diagram extends up to ~14 at. % Cr, while in Fe-Ni the fcc structure is more stable than bcc for almost all the concentrations and temperatures. This is the consequence of strong ferromagnetic interaction between iron and nickel in fcc phase resulting in large negative energies of mixing seen in *ab initio* calculations. Our parameterization of MCE Hamiltonian also predicts strong ferromagnetic exchange coupling between the second and third nearest Fe-Ni neighbours and negative enthalpy of mixing, in excess of -100 meV/atom, even for random alloy configurations. Magnetic behaviour of random alloy configurations is characterized by the transition from non-collinear antiferromagnetism on the iron-rich side to collinear ferromagnetism after the addition of relatively moderate (~25 at. %) amount of nickel. Large, compared to Ni, magnetic moment of Fe, results in a maximum of the total magnetic moment of the system as a function of concentration. At elevated temperatures, magnetic order vanishes first on the Fe-rich side of the concentration range.

An important feature of FeNi alloy system is the occurrence of several ferromagnetically ordered intermetallic compounds. While the magnetic order in $L1_2$ superlattice of $Fe_3Ni$ disappears at relatively low temperatures, FeNi and $FeNi_3$ remain ferromagnetic even above 1000 K. In fact, this is the main reason why they become unstable with respect to transformation into chemically disordered configurations. The Curie temperatures for random mixtures are much lower than for the ordered structures, and above the corresponding Curie temperatures the disordered configurations are paramagnetic. It is well known that in the paramagnetic state the magnetic energy increases with temperature much slower than in a ferromagnetic state (Figure 11), hence at higher temperature the ordered compounds become less energetically favourable than random alloy mixtures. It is worth noting here that one previous theoretical study of order-disorder phenomena in $Fe_3Ni$, FeNi, and $FeNi_3$ [38] found magnetic order even in chemically disordered alloys at temperatures where they are more stable than the ordered ones. The study was based on an Ising Hamiltonian, which appears unrealistic for the fairly magnetically isotropic Fe-Ni compound. Our calculations also show that in the limit of a strongly magnetically anisotropic MCE Hamiltonian, chemically disordered system remain ferromagnetic until higher temperatures. However, experimental study [44] found only small magnetic crystal anisotropy in iron–nickel alloys. We believe that a realistic Hamiltonian for that system should be almost entirely isotropic, and some magnetic order above the chemical order-disorder transition may be related to precipitates of ordered phases or to short-range chemical order remaining at temperatures where long-range chemical order is already absent.

For the Z1 $Fe_3Ni$ superlattice, non-collinear antiferromagnetic order is predicted. It is reasonable to expect that other antiferromagnetic ordered structures may also exist in the region of high Fe concentration, with their magnetic structures similar to that of pure fcc iron. The energies of Z1 and L1$_2$ structures in Fe$_3$Ni are close to each other, in accord with MCE simulations. Also, we note that our DFT calculations performed for several different structures with 25 at. % Ni content give fairly similar energies. These structures can be both completely ferromagnetic or partially antiferromagnetic (ferrimagnetic), with some of the Fe atoms having magnetic moments directed opposite to the rest of the moments, without involving large penalty in energy. The volumes of these structures as calculated by DFT differ by 2-3 % (with ferrimagnetically ordered systems having smaller volume) – an effect which cannot be modelled using rigid lattice MCE simulations. This abundance of superlattices and magnetic structures within a narrow energy range is in agreement with previous calculations [45] that relate the occurrence of the ferrimagnetic phase to the Invar effect at around 35 at. % Ni. Our MCE calculations exploring several superlattices with Fe$_2$Ni composition found both ferromagnetic as well as ferrimagnetic ground states, but the antiferromagnetic configuration seems to be too high in energy for the specific Ni concentration and cannot be obtained without constraining the total magnetic moment of the system. Another possible reason explaining the Invar effect, the non-collinearity of magnetic structures [46], has also been observed in our simulations for both the ordered (Z1) and random Fe-Ni mixtures with Ni content not exceeding that of Fe.

The model developed above is limited to classic Heisenberg-Landau-like Hamiltonians. Recent studies of pure Fe and Cr [47,48] showed the significance of taking quantum corrections and lattice anharmonicity into account at high temperatures, to achieve fully quantitative description of the phase diagram. Still, the methodology proposed in [47,48] is yet to be extended to solid solutions of metals. In our previous work [9], by taking into account vibrational contributions to the free energy we correctly estimated the size of the γ-loop in Fe-Cr, but were only able to do that in a narrow range of Cr concentrations not exceeding 15 at. % Cr. For the fcc Fe-Ni alloy studied here, the vibrational contribution to the free energy appears to be very small [39], and the current parameterization of the MCE Hamiltonian without quantum and vibrational corrections made it possible to undertake realistic study of magnetic and thermodynamic properties of the alloy in a broad range of temperatures and concentrations. Good agreement with experiment was found for the fcc-bcc coexistence curve in the phase diagram and for the temperatures of order-disorder transitions in FeNi, and FeNi$_3$ compounds. This shows that the MCE model can now be extended to the ternary Fe-Ni-Cr alloy.

**Acknowledgement**

This work, part-funded by the European Communities under the contract of Association between EURATOM and CCFE, was carried out within the framework of the European Fusion Development Agreement. To obtain further information on the data and models underlying this paper please contact *PublicationManager@ccfe.ac.uk*. The views and opinions expressed herein do not necessarily reflect those of the European Commission. This work was also part-funded by the RCUK Energy Programme under grant EP/I501045. DNM would like to thank Juelich supercomputer centre for using the High-Performances Computer for Fusion (HPC-FF) facilities as well as the International Fusion Energy Research Centre (IFERC) for using the supercomputer (Helios) at Computational Simulation Centre (CSC) in Rokkasho (Japan). JSW is supported by the Accelerated Metallurgy Project, which is co-funded by the European Commission in the 7th Framework Programme (Contract NMP4-LA-2011-263206), by the European Space Agency and by the individual partner organisations.

**References**

1. S.L. Dudarev *et al.*, *J. Nucl. Mater.* **386-388**, 1 (2009).

2. T. Toyama *et al.*, *J. Nucl. Mater.* **425**, 71 (2012).

3. D. Terentyev and A. Bakaev, *J. Nucl. Mater.* **442**, 208 (2013).

4. R. Meyer and P. Entel, *Phys. Rev. B* **57**, 5140 (1998).

5. Y. Mishin, M.J. Mehl, and D.A. Papaconstantopoulos, *Acta Mater.* **53**, 4029 (2005).

6. G. Bonny, R.C. Pasianot, and L. Malerba, *Modelling Simul. Mater. Sci. Eng.* **17,** 025010 (2009).

7. G. Bonny, R.C. Pasianot, and L. Malerba, *Phil. Mag.* **89**, 3451 (2009).

8. M.Yu. Lavrentiev, S.L. Dudarev, and D. Nguyen-Manh, *J. Nucl. Mater.* **386-388**, 22 (2009).

9. M.Yu. Lavrentiev, D. Nguyen-Manh, and S.L. Dudarev, *Phys. Rev. B* **81**, 184202 (2010).

10. M.Yu. Lavrentiev, D. Nguyen-Manh, and S.L. Dudarev, *Comp. Mat. Sci.* **49**, S199 (2010).

11. M.Yu. Lavrentiev, S.L. Dudarev, and D. Nguyen-Manh, *J. Appl. Phys.* **109**, 07E123 (2011).

12. V.I.L.J. Swartzendruber and C. Alcock, Phase Diagrams of Binary Iron Alloys (Materials Park, OH: ASM International, 1993).

13. J.F. Albertsen, G.B. Jensen, and J.M. Knudsen, *Nature* **273**, 453 (1978).

14. R.S. Clarke, Jr. and E.R.D. Scott, *American Mineralogist* **65**, 624 (1980).

15. C.-W. Yang, D.B. Williams, and J.I. Goldstein, *Geochimica et Cosmochimica Acta* **61**, 2943 (1997).

16. M.Yu. Lavrentiev, D. Nguyen-Manh, and S.L. Dudarev, *Solid State Phenomena* **172-174**, 1002 (2011).

17. Y. Zhang et al., *Progress in Materials Science* **61**, 1 (2014).

18. P.E. Blöchl, *Phys. Rev. B* **50**, 17953 (1994).

19. G. Kresse, J. Hafner, *Phys. Rev. B* **49**, 14251 (1994).

20. G. Kresse, J. Furthmüller, *Comp. Mater. Sci.* **6**, 15 (1996).

21. G. Kresse, J. Furthmüller, *Phys. Rev. B* **54**, 11169 (1996).

22. J.P. Perdew, K. Burke, M. Ernzerhof, *Phys. Rev. Lett.* **77**, 3865 (1996).

23. H. Wang et al., *Phys. Chem. Chem. Phys.* **15**, 7599 (2013).

24. M.Yu. Lavrentiev et al., *J. Phys. Chem. B* **105**, 3594 (2001).

25. F.M. Marquez et al., *Modelling Simul. Mater. Sci. Eng.* **11**, 115 (2003).

26. M.Yu. Lavrentiev, J.A. Purton, and N.L. Allan, *American Mineralogist* **88**, 1522 (2003).

27. M.Yu. Lavrentiev, N.L. Allan , and J.A. Purton, *Phys. Chem. Chem. Phys.* **5**, 2190 (2003).

28. M.Yu. Lavrentiev et al., *Phys. Rev. B* **75**, 014208 (2007).

29. Y. Tsunoda, *J. Phys.: Condensed Matter* **1**, 10427 (1989).

30. A. Onodera et al., *Phys. Rev. B* **50**, 3532 (1994).

31. M.Yu. Lavrentiev et al., *Phys. Rev. B* **84**, 144203 (2011).

32. Q. Chen and B. Sundman, *J. Phase Equilibria* **22**, 631 (2001).

33. S. Blundell, *Magnetism in Condensed Matter*, Oxford University Press, 2001.

34. R. Idczak, R. Konieczny, and J. Chojcan, *Physica B* **407**, 235 (2012).

35. R.J. Wakelin and E.L. Yates, *Proc. Phys. Soc. London Sect B* **66**, 221 (1953).

36. T. Mohri and Y. Chen, *J. Alloys and Compounds* **383**, 23 (2004).

37. S.V. Barabash et al., *Phys. Rev. B* **80**, 220201 (2009).

38. M.Z. Dang and D.G. Rancourt, *Phys. Rev. B* **53**, 2291 (1996).

39. P.D. Bogdanoff and B. Fultz, *Phil. Mag. B* **79**, 753 (1999).

40. J. Paulevé et al., *C. R. Acad. Sci.* **254**, 965 (1962).

41. J.W. Drijver, F. van der Woude, and S. Radelaar, *Phys. Rev. Lett*. **34**, 1026 (1975).

42. I. Leonov et al., *Phys. Rev. Lett.* **106**, 106405 (2011).

43. F. Körmann et al., *Phys. Stat. Sol. B* **251**, 53 (2014).

44. R.M. Bozorth and J.G. Walker, *Phys. Rev.* **89**, 624 (1953).

45. I.A. Abrikosov et al., *Phys. Rev B.* **76**, 014434 (2007).

46. M. van Schilfgaarde, I.A. Abrikosov, and B. Johansson, *Nature* **400**, 46 (1999).

47. F. Körmann et al., *Phys. Rev B.* **81**, 134425 (2010).

48. F. Körmann et al., *J. Phys.: Condens. Matter* **25**, 425401 (2013).

**Tables**

|  | Fe-Fe | Fe-Ni | Ni-Ni |
|---|---|---|---|
| 1st nearest neighbour | -0.793 | 1.516 | -13.153 |
| 2nd nearest neighbour | -10.827 | -2.710 | 7.228 |
| 3rd nearest neighbour | 0.547 | -2.500 | -5.605 |
| 4th nearest neighbour | 2.306 | 1.649 | -6.744 |

Table 1. Magnetic Heisenberg interaction parameters $J_{ij}$ (in meV units) fitted to *ab initio* data and used in the MCE simulations described below.

|  | MCE | DFT |
|---|---|---|
| Non-magnetic | 0 | 0 |
| Low-spin ferromagnetic | -12 | -5 |
| High-spin ferromagnetic | -19 | -14 |
| Single layer antiferromagnetic | -51 | -39 |
| Double layer antiferromagnetic | -51 | -60 |
| Non-collinear antiferromagnetic | -53 |  |

Table 2. Energies of several magnetic configurations of fcc iron (meV/atom) found in MCE and DFT calculations. The values of the energies are given with respect to the energy of a non-magnetic state.

**Figures**

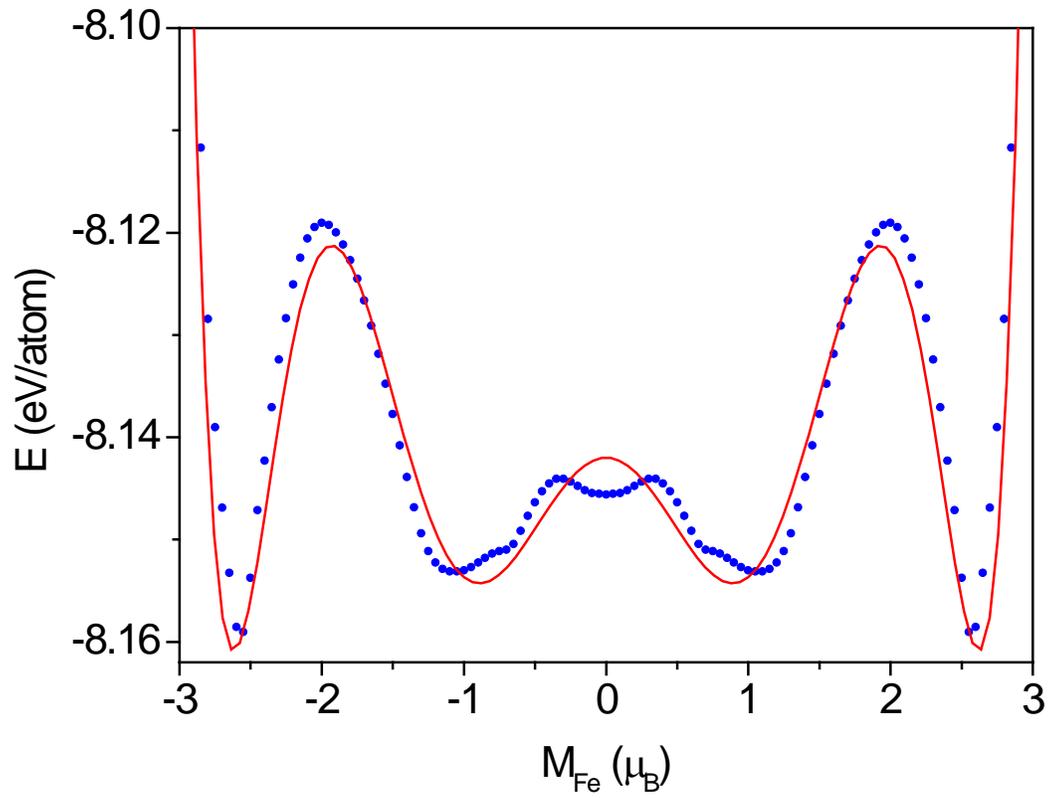

Figure 1. Comparison of DFT (points) energies of ferromagnetic fcc Fe plotted as a function of magnetic moment, and the MCE Landau expansion (line), which includes terms up to the 8th order in atomic magnetic moment.

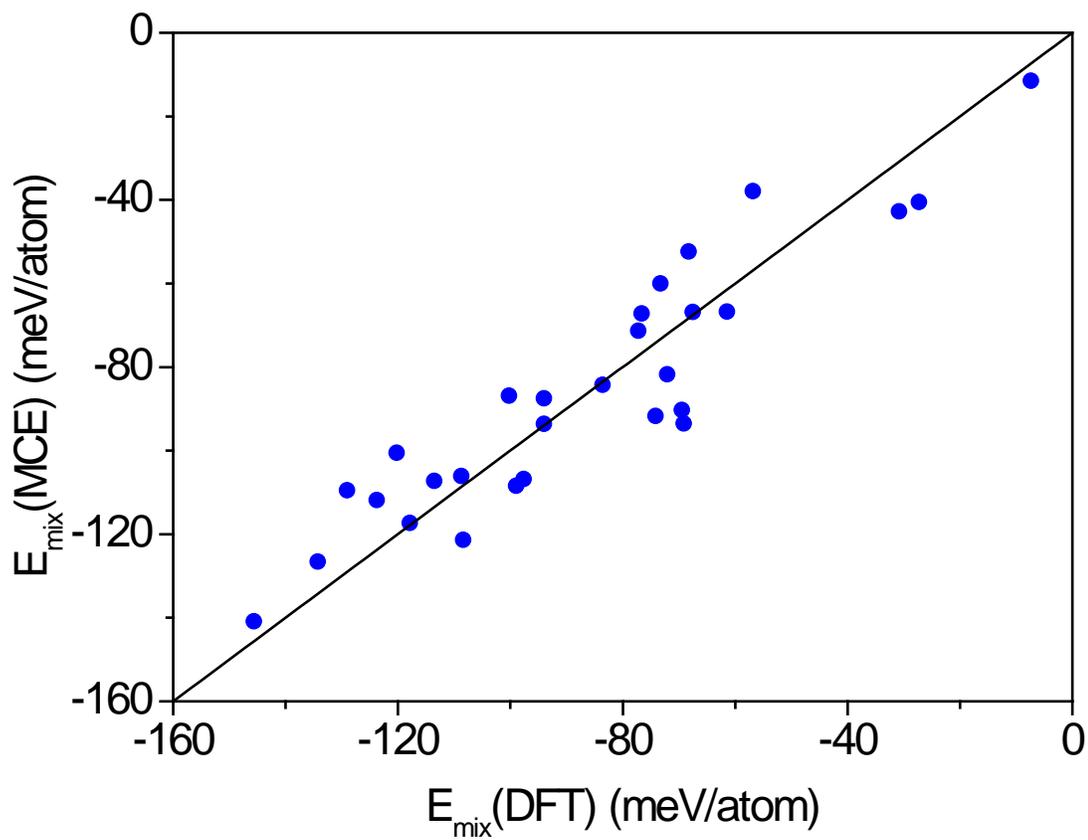

Figure 2. Comparison of DFT and MCE energies of mixing for the alloy configurations used for fitting the MCE Hamultonian.

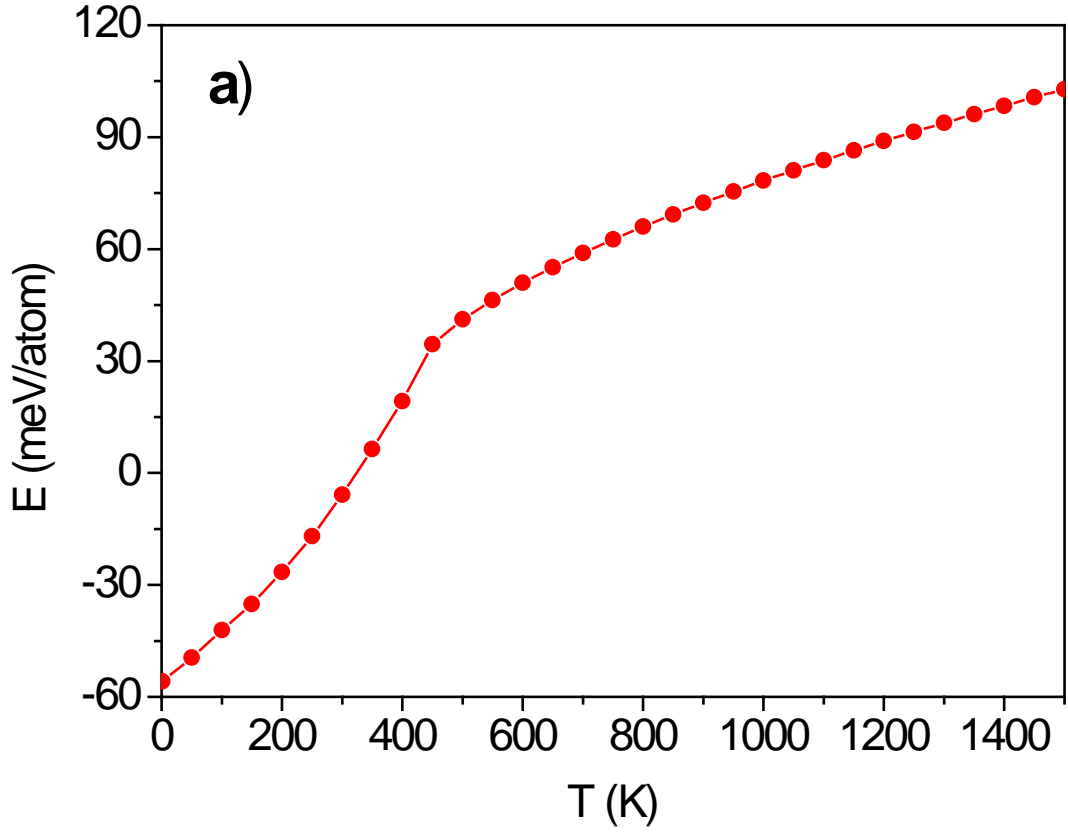

Figure 3a. Energy of pure fcc iron (in meV/atom units) plotted as a function of temperature.

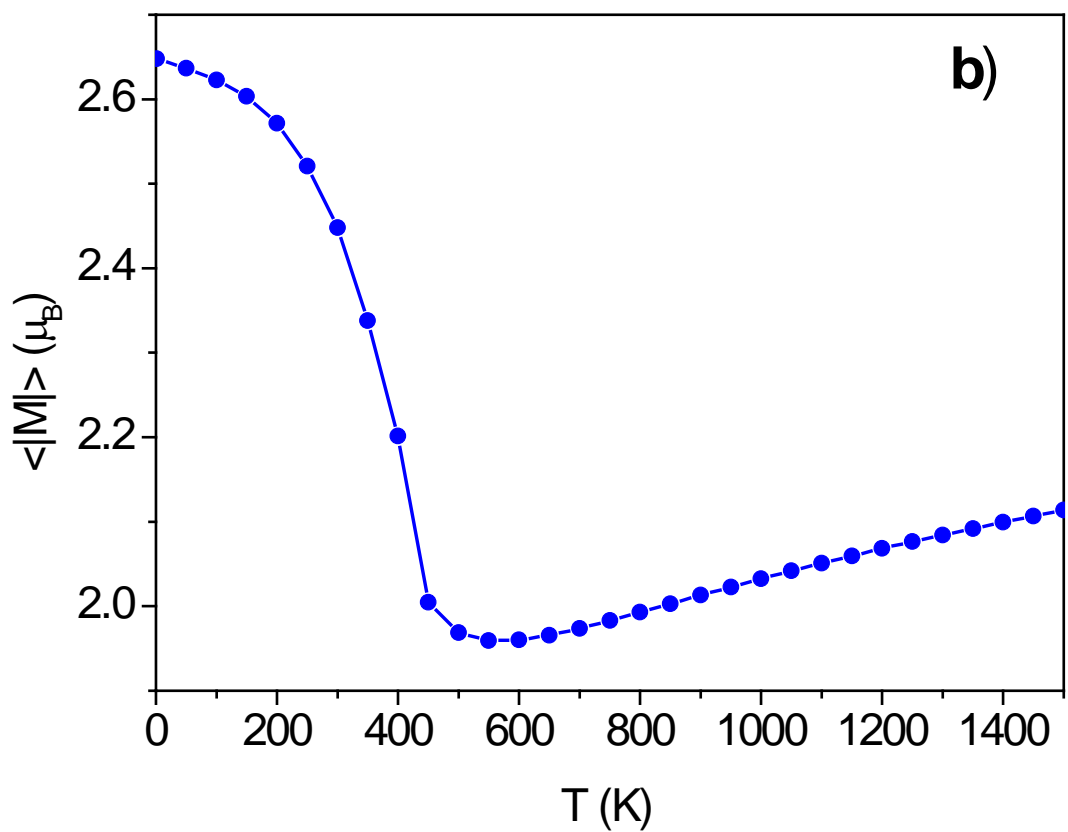

Figure 3b. Average length of atomic magnetic moment in fcc Fe ($\mu_B$) plotted as a function of temperature.

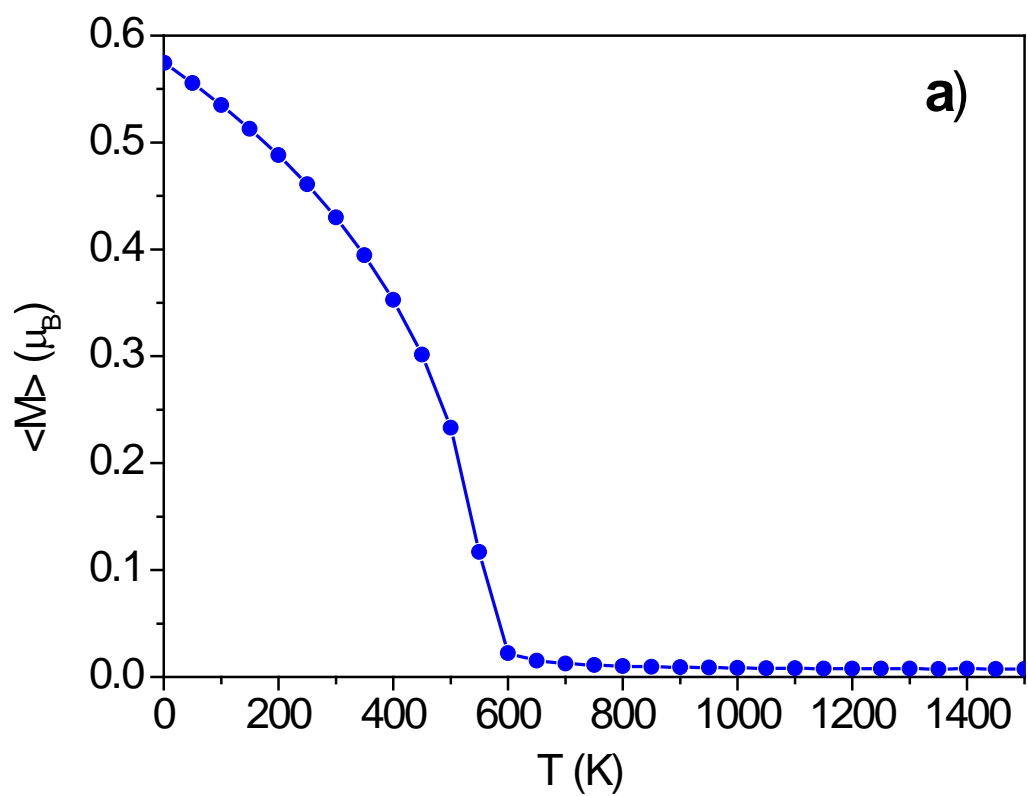

Figure 4a. Magnetic moment of pure nickel ($\mu_B$) plotted as a function of temperature.

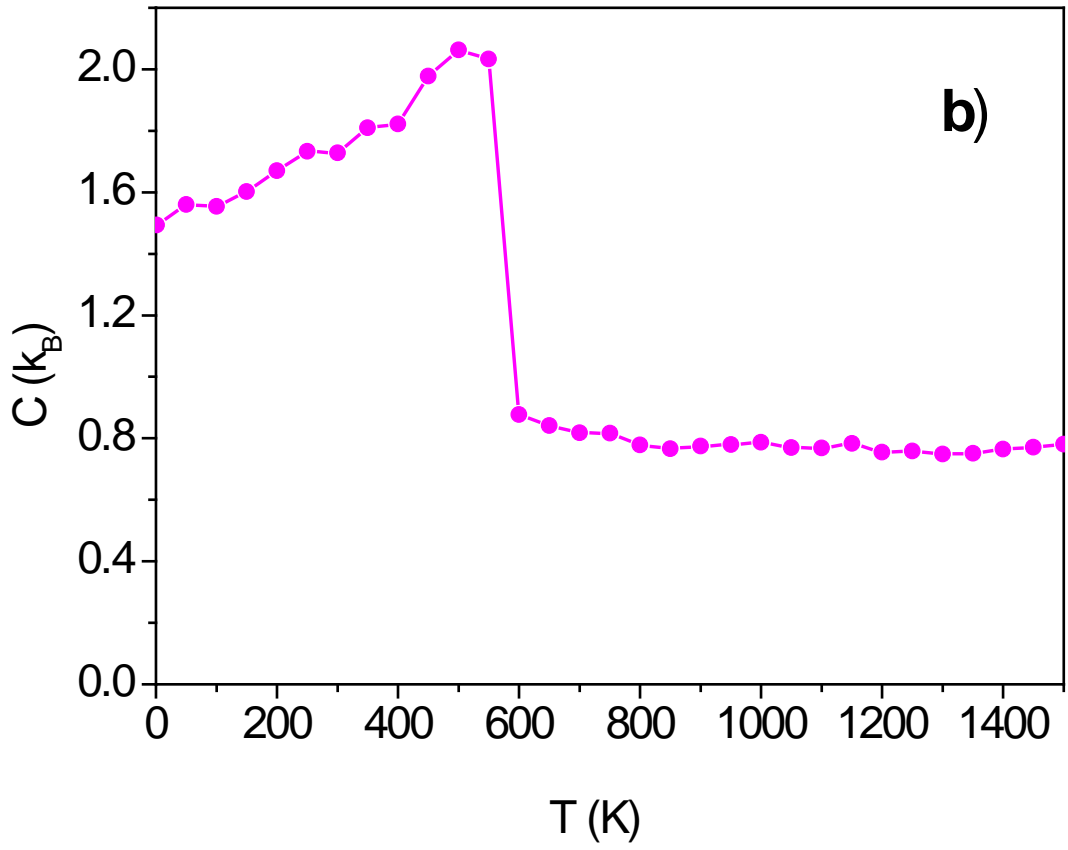

Figure 4b. Temperature dependence of the magnetic part of the specific heat computed using MCE for pure nickel.

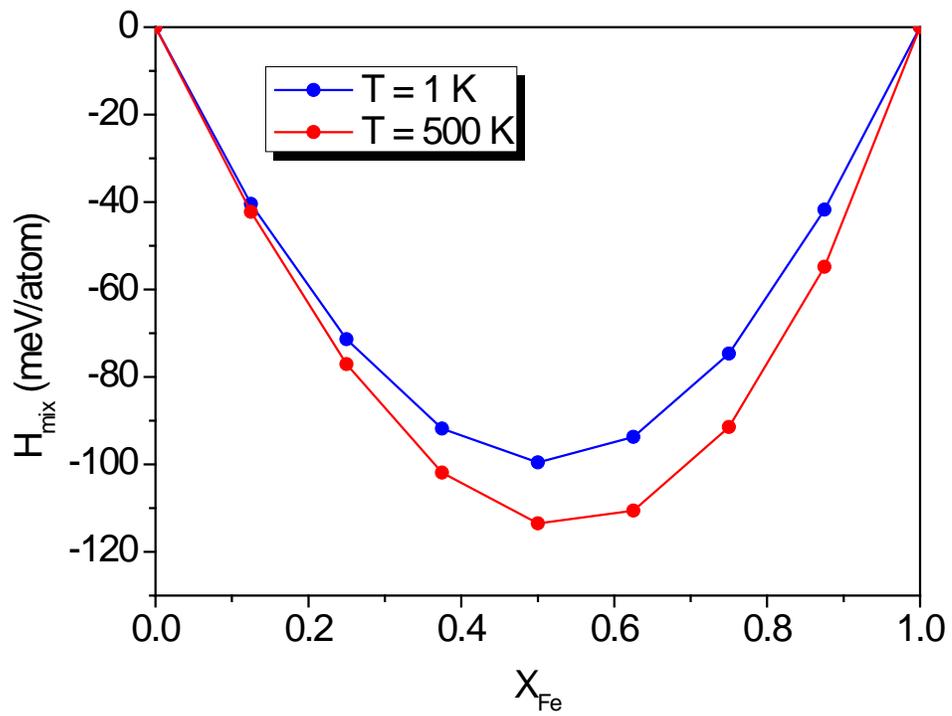

Figure 5. Enthalpy of mixing of random fcc Fe-Ni alloy configurations computed for low and high temperatures T=1 K and T=500 K.

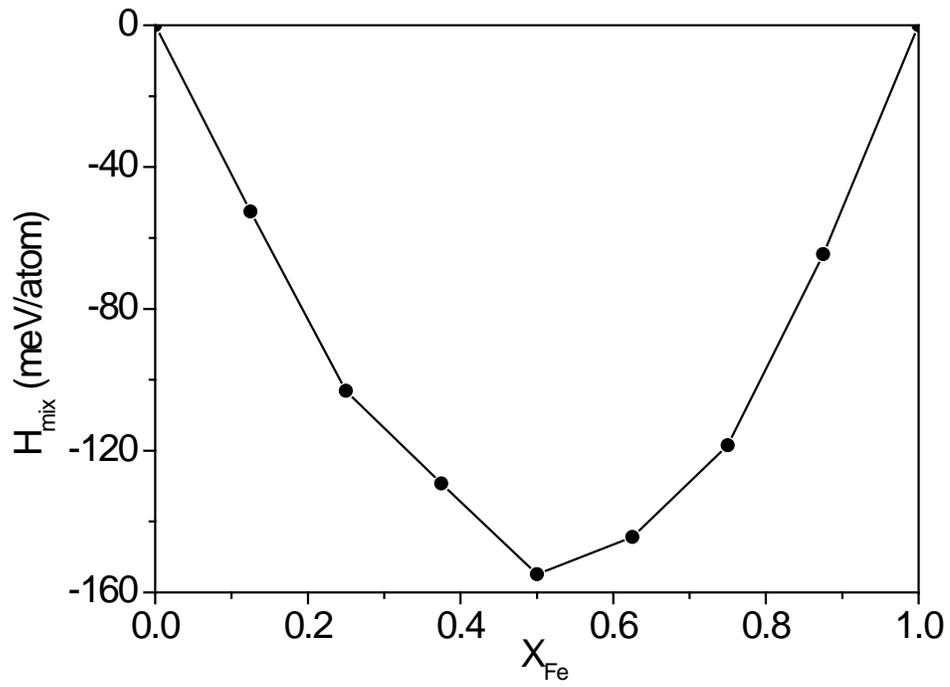

Figure 6. Enthalpy of mixing of FeNi alloys plotted as a function of Fe content for quenched configurations.

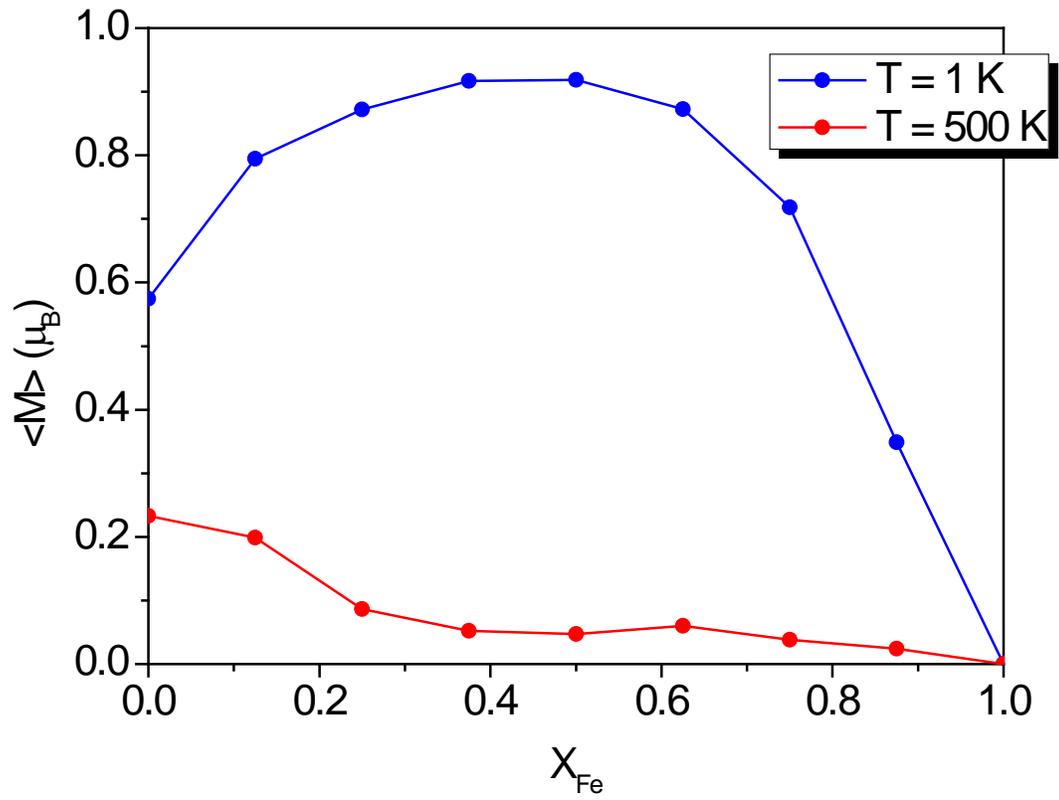

Figure 7. The total magnetic moment of a random Fe-Ni mixture plotted as a function of Fe content at a very low temperature and at T = 500 K.

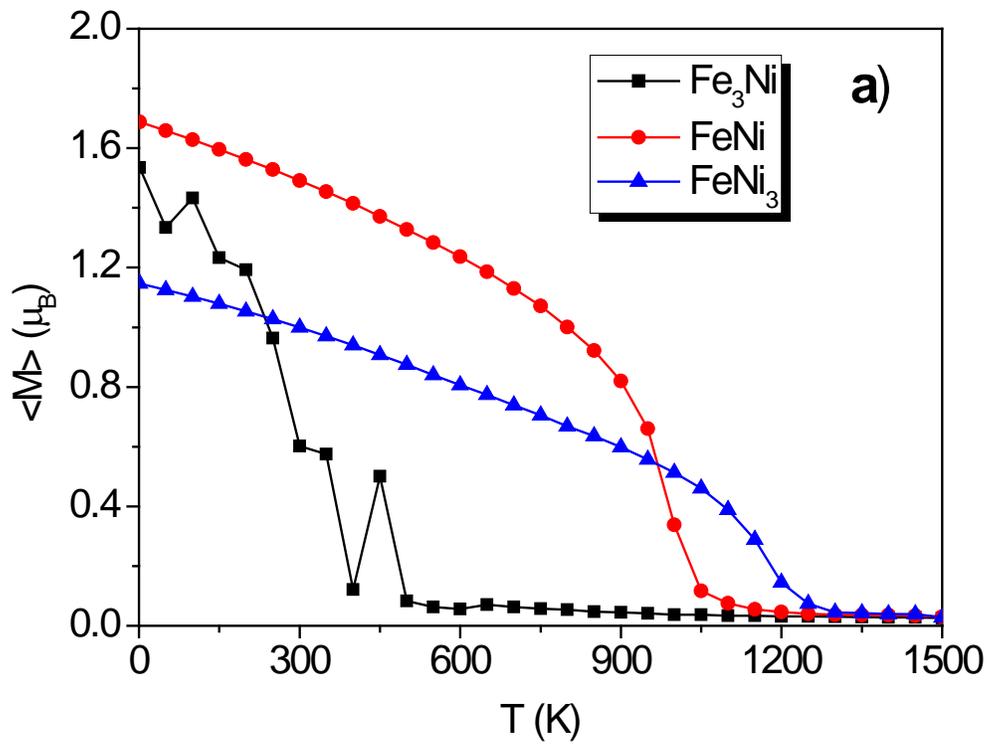

Figure 8a. Temperature dependence of magnetic moments predicted for several ordered Fe-Ni compounds.

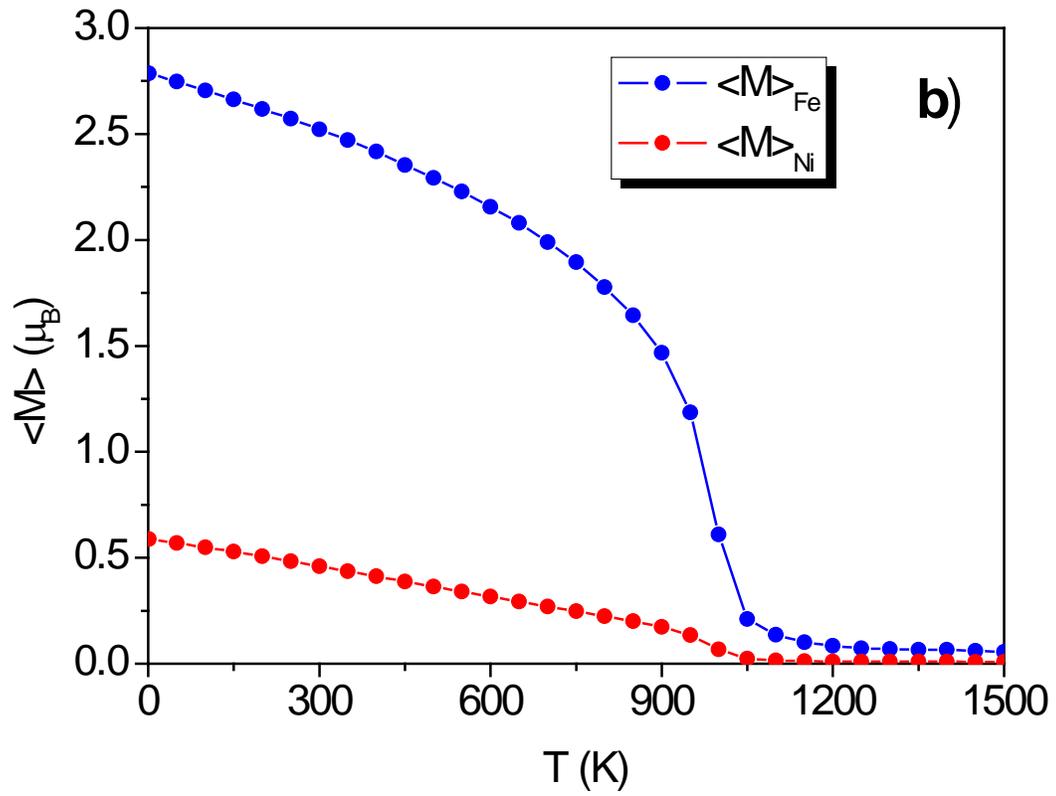

Figure 8b. Magnetic moments of Fe and Ni sublattices in an ordered FeNi alloy plotted as a function of temperature.

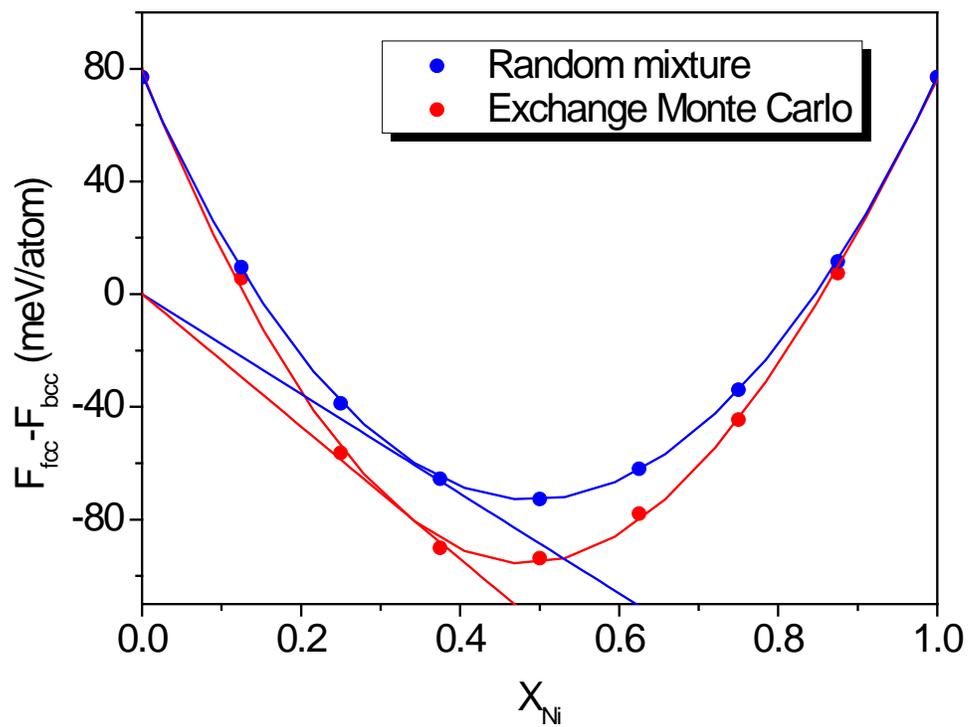

Figure 9. Tangent construction defining the coexistence curve between bcc Fe and fcc Fe-Ni alloys at temperature T=600 K.

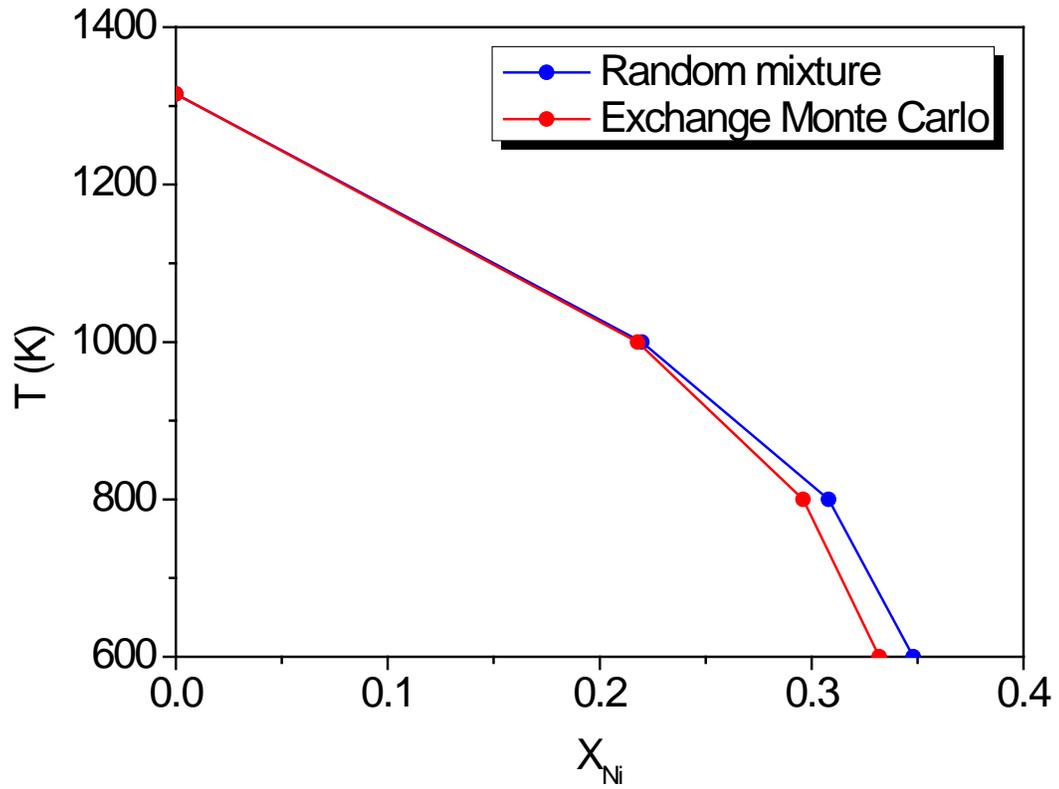

Figure 10. Coexistence curves between bcc Fe and fcc Fe-Ni alloys. The blue line corresponds to random alloys, the red line was computed using exchange Monte Carlo simulations.

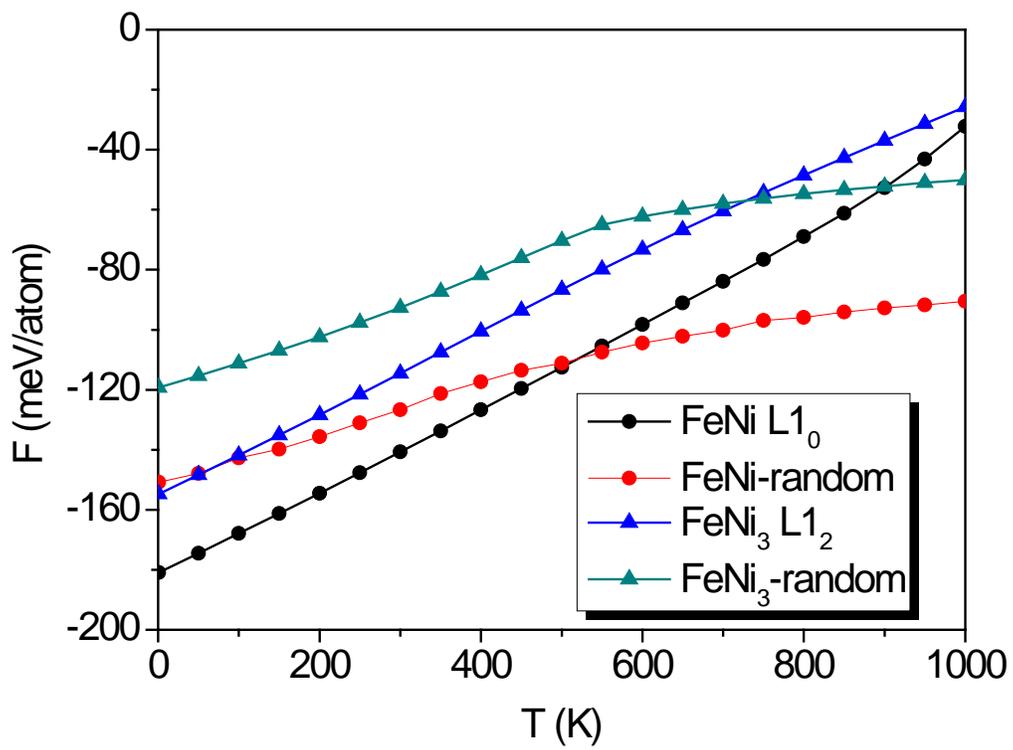

Figure 11. Comparison of free energies of ordered and random structures simulated using MCE for stoichiometric compositions FeNi and FeNi$_3$.